# The Algorithmic Imprint

Upol Ehsan
Georgia Institute of Technology
ehsanu@gatech.edu

Ranjit Singh
Data & Society Research Institute
ranjit@datasociety.net

Jacob Metcalf
Data & Society Research Institute
jake.metcalf@datasociety.net

Mark O. Riedl
Georgia Institute of Technology
riedl@cc.gatech.edu

## ABSTRACT

When algorithmic harms emerge, a reasonable response is to stop using the algorithm to resolve concerns related to fairness, accountability, transparency, and ethics (FATE). However, just because an algorithm is removed does not imply its FATE-related issues cease to exist. In this paper, we introduce the notion of the "algorithmic imprint" to illustrate how merely removing an algorithm does not necessarily undo or mitigate its consequences. We operationalize this concept and its implications through the 2020 events surrounding the algorithmic grading of the General Certificate of Education (GCE) Advanced (A) Level exams, an internationally recognized UK-based high school diploma exam administered in over 160 countries. While the algorithmic standardization was ultimately removed due to global protests, we show how the removal failed to undo the algorithmic imprint on the sociotechnical infrastructures that shape students', teachers', and parents' lives. These events provide a rare chance to analyze the state of the world both with and without algorithmic mediation. We situate our case study in Bangladesh to illustrate how algorithms made in the Global North disproportionately impact stakeholders in the Global South. Chronicling more than a year-long community engagement consisting of 47 interviews, we present the first coherent timeline of "what" happened in Bangladesh, contextualizing "why" and "how" they happened through the lenses of the algorithmic imprint and situated algorithmic fairness. Analyzing these events, we highlight how the contours of the algorithmic imprints can be inferred at the infrastructural, social, and individual levels. We share conceptual and practical implications around how imprint-awareness can (a) broaden the boundaries of how we think about algorithmic impact, (b) inform how we design algorithms, and (c) guide us in AI governance. The imprint-aware design mindset can make the algorithmic development process more human-centered and sociotechnically-informed.

## 1 INTRODUCTION

> *"Aah now I see it! Even though they took the algorithmic calculation out, its ghost [spirit] lived on in the revised output"—A participant reflecting on the outcome of a revised algorithmic grading event*

A common—and reasonable–response upon discovery of harms from an algorithmic system is to simply stop using it so that its harmful effects are prevented from propagating farther. This has been a common approach to address public demands for accountability of harmful systems; for instance, the Gender Shades studies [16, 17] highlighted the harms of biased commercial facial recognition systems, catalyzing their discontinuation [3, 26]. While ceasing algorithmic use is a good first step, such an approach may be insufficient for mitigating all such harms. We elucidate why that is the case through the events surrounding the algorithmic grading of the globally administered GCE A Level exams in June 2020.

In this paper, we introduce the notion of the *algorithmic imprint* to illustrate how the removal of an algorithm does not necessarily entail that its effects are stopped or undone. Put differently, algorithmic consequences extend well beyond the algorithm's "lifetime" (deployment period). The concept of *algorithmic imprint* broadens the boundaries of what is typically considered for algorithmic impact. Typically, our conceptual canvas of algorithmic fairness issues is most salient in and circumscribed to the algorithm's lifetime [86]. Like a footprint remains the sand long after someone has passed, or a palimpsest records the writing on the sheet of paper above it, algorithmic systems can leave their mark on the data infrastructure, societal organization, or mental wellbeing of data subjects long after stopping their use. The concept of the imprint does not change the existing nature of algorithms. Algorithms have always been imprint-laden. The notion of the imprint provides the essential vocabulary to articulate and address harms and ethical issues in the algorithm's "afterlife" (period after deployment ceases)—an unexplored area in the algorithmic fairness literature. Algorithmic harms can be harder to detect in the afterlife because the algorithm's absence can make these issues less visible. The algorithmic imprint underscores that *algorithms do not function in a vacuum;* a wide range of infrastructural conditions and organizational practices (that make algorithms possible) act in concert to co-constitute stakeholders' lived experiences of the algorithm's consequences [28, 76]. The imprint forms a junction that connects the consequential infrastructural conditionals and organizational practices that create and sustain the deployment during algorithm's lifetime with their after-effects that persist in the afterlife. It provides an important point of inquiry to investigate the differential agencies and experiences of the stakeholders for possible algorithmic harms.

Furthermore, just because imprints can persist in the algorithmic afterlife does not mean they are not present during the lifetime. The imprints can develop and evolve throughout a deployment. The contours of imprints emerge and can be detected at many levels: at the infrastructural level, they can develop through formative data practices that make the algorithm possible. At the social level, they can be discerned through the altered social relationships (how people relate to each other) due to algorithmic mediations, constraining how a society addresses collective problems. At an individual level, algorithms can leave imprints on how people make sense of algorithmic operations and interpret their lived experiences with the algorithm, carrying deep psychological impact on their mental well-being. An imprint-aware approach can afford the ability to "see" effects that would otherwise be less visible due to the absence of the algorithm. By making algorithmic fairness issues traceable and tractable in the algorithm's afterlife, the imprint can bring the oft-invisible elements (e.g., data practices) to conscious awareness, thereby making them actionable. Serving as a junction point connecting the algorithmic lifetime with its afterlife, the concept can allow us to trace less visible infrastructural facets to find areas of harm or missing accountability.

We situate the concept through the algorithmic grading events of the General Certificate of Education (GCE) Advanced (A) Level exams in 2020. GCE A Levels are globally-recognized international exams that evaluate high school students' academic competence [115]. Even though the exam boards are UK-based [81], these exams are taken in more than 160 countries, many of them are a part of the British Commonwealth (an association of nations that were formerly British colonies). The A Level grades are consequential, serving an indispensable role in university admissions. Due to the COVID-19 pandemic, the Office of Qualifications and Exam Regulations (aka 'Ofqual')—the UK-based quasi-governmental office that oversees GCE exams—cancelled in-person exams. In lieu of actual exams, they used teacher assessments of school records to generate an algorithmically-normed grade. In August 2020, global protests erupted after the release of the algorithmic grades, citing inconsistencies and socio-statistical biases [4, 113]. The exam boards ultimately reversed course [84]—they globally rescinded the algorithmic-assessed grades and revised the grades just based on teacher assessed grades without the algorithmic standardization, implying removal of the algorithm's effect [82]. However, as we illustrate in this paper, taking the algorithmic calculation out did not mean that the investments in the processes to achieve algorithmic grading were also undone. *Like the remnant traces on a palimpsest, the imprint remained even after the algorithm was removed.* These events provide a rare opportunity to witness the afterlife of a globally deployed algorithm and to *investigate the state of the world with and without the algorithm* explicitly at play.

To situate the imprint in a practical setting, we chose Bangladesh as the site for our case study conducting more than a year-long community engagement consisting of 47 interviews and over 100 informal conversations. Instead of the UK events (which received considerable press attention [47, 78]), we chose Bangladesh for two reasons: first, there was a dearth of press coverage of the consequences of Ofqual's decisions and student protests outside of the UK, despite the fact that commonwealth students and teachers constitute significant proportions of stakeholders for GCE exams [116]. Second, empirical insights from under-explored geographies and contexts in "the Global South" can potentially highlight novel ways of living with and navigating algorithmic interventions developed in "the Global North" for FAccT and allied fields. Algorithmic systems are often designed with particular infrastructural assumptions of places where they are designed (or, centers of production) around available data and logistical support [24, 62]. Focusing on proverbial geographic peripheries that do not confirm with these assumptions allows us to map core struggles of operationalizing algorithmic systems at scale and grappling with the challenging consequences of their deployment.

Below, we ground the algorithmic imprint through related work in situated algorithmic fairness and folk theories. Next, we share our methods and contextualize the examinations through Bangladeshi perspectives. Then, we construct a coherent timeline of "what" happened in Bangladesh, contextualizing "why" and "how" they happened through the lenses of the algorithmic imprint and conclude by sharing implications. In summary, our contributions are three-fold.

- We introduce the concept of the *algorithmic imprint* to illustrate (1) how an algorithm's impact extends beyond its lifetime into its afterlife and (2) how merely removing an algorithm does not necessarily undo its consequences.
- Utilizing the insights from 47 interviews, we present the first coherent timeline that chronicles the Bangladeshi events around the 2020 algorithmic grading of GCE A level exams and situate the events through the lenses of the imprint
- We discuss implications at the design and governance levels, highlighting how imprint-awareness can inform how developers design algorithms and how AI governance measures facilitate it

## 2 GROUNDING THE ALGORITHMIC IMPRINT

We conceptualized the algorithmic imprint—wherein the impacts of an algorithmic system persist beyond its use—by drawing on three distinct yet related threads: 1) infrastructure studies, 2) situated fairness, and 3) folk theories and perceptions around the workings of algorithms. Weaving these threads together, we chart the otherwise invisible underlying infrastructural conditions and practices that make algorithmic deployments possible and contribute to their potentially harmful impacts [103, 104]. Our work is oriented to describing the boundaries of an algorithmic system, and by extension, its constitution; it draws on and contributes to studies of the sociopolitical dynamics of algorithmic systems that trace the boundaries of a system by its effects rather than by its technical specifications [42, 91]. The methodological question of how to bound algorithmic systems has significant consequences for questions of algorithmic accountability.

## 2.1 Infrastructural Lens

Broadly, the body of algorithmic fairness literature consists of two distinct, complementary analytic lenses to elucidate the ethical and social consequences of algorithmic systems: *the technical lens* and *the contextual lens*. We use these lenses as shorthands to guide *how the contours of the algorithm's consequences define the system*: when a harm is identified, where do we look to find its source? These lenses are neither mutually exclusive nor fully self-contained. We use them to illustrate what is distinct about algorithmic imprint and what we identify as a third, *the infrastructural lens*.

*The technical lens* finds the ethical consequences of algorithmic systems in the methods and technical details of how the systems are constructed, focusing interventions on best practices for constructing, modifying, or governing algorithmic systems. Exemplars of the technical perspective include developing technical documentation and organizational accountability practices [39, 73, 86], audits of the outputs of algorithmic systems [17, 80], demonstrations of the perils of large datasets [11, 48, 56], and formalized models of algorithmic fairness [10, 77]. With the technical lens, the algorithmic system is bounded by the statistical and material components that are responsible for the harm—if one wants to understand the algorithmic system and its consequences, one looks to how it is technically built and operated.

*The contextual lens* finds the algorithmic ethical consequences in the social and political relations that the system interfaces with. Exemplars includes accounts of organizational roles and values in AI/ML ethics [1, 5, 69], studies of how data technologies redistribute social power [35, 85, 93], how algorithmic systems can reinforce extant social and political biases [79, 90], and recapitulate a worldview with harmful and undesirable histories [7, 13, 71]. With the contextual lens, the algorithmic system is bounded by (a) the social context that has been formalized as training data and (b) the context of the deployed site—if one wants to understand the algorithmic system and its consequences, one looks to its context.

*The infrastructural lens* utilized here instead examines the social and political conditions *that make the system possible in the first place*. It asks: how must human lives and social systems be structured for this algorithmic system to operate and continue functioning? The focus is on the *uneven imbrication of algorithmic systems* [64] *with existing sociotechnical arrangements* of people and things that form a data infrastructure [44, 73]. Data infrastructures involve connecting diverse existing practices that do not seamlessly stack on top of each other [25, 101, 107]. When this infrastructure works as expected, the practices holding the imbrication together become invisible. They are made visible when the imbrication breaks down [40, 104]. Grounded in the insight that infrastructures work for some at the expense of others [27], the infrastructural lens focuses either on the unevenness of the imbrication [24, 63, 83, 100] or on such breakdowns and harmful consequences [33, 41, 59, 104].

This lens emphasizes the agency and responsibility of analysts in making choices—or what Suchman has called making "accountable cuts" [106]—around which conditions and processes to focus on and which lived experiences of an infrastructure deserve attention [38, 102]. With the infrastructural lens, the algorithmic system is bounded by its imbrication with existing conditions and processes.

## 2.2 Situated Fairness

From the infrastructural perspective, an imprint is not just technical components of a system left behind after the algorithm is no longer in use, but rather an account of its impacts as experienced by people who are subject to it. How should we come to understand those impacts to form a more coherent view of the imprint from slices of infrastructure? Here we turn to *situatedness* (aka *situated knowledges* and *situated methods*), a notion that emerges foremost from feminist Science & Technology Studies (STS). Haraway coined "situated knowledges" [45] to describe how it is impossible to achieve objectivity as a God's-eye-view or a "view from nowhere" of a technoscientific system; she argues instead for an epistemic and ethical commitment to embodied and accountable objectivity wherein the analyst is responsible for situating themselves and others in producing an account how a system is constituted and what are its consequences. The emphasis on partial, embodied knowledge in situated methods also provides a model for understanding how algorithmic fairness can be grounded in the experiences of those most impacted by algorithmic systems through participatory research [53], guidelines for collecting research data [21] and revising target states in statistical measures of fairness [36]. Indeed, strictly abstract and technical accounts of algorithmic un/fairness often fail to capture the full range of potential algorithmic harms and may provide cover for deeper issues of injustice [44, 74, 77, 90, 92].

Building on this line of research, we propose an approach to *situated fairness* where descriptions of how an algorithmic system is constituted and its consequences are grounded in accounts of people who live with it. Viljeon [108] argues that machine learning is fundamentally a matter of relating individuals—their grades, their credit, etc.—to a collective of historical data, through mechanisms of statistical prediction. Similarly, Jacobs & Wallach [50] argue that many contestations about algorithmic fairness are actually conflicts over the validity of inferring unobservable theoretical constructs (e.g., worker effectiveness) from observable features that are readily available in datasets. *Situated fairness is a matter of asking about the lived conditions of that relating—* what is the lived experience of a predictive mechanism that relates individuals to collective(s)? What infrastructures are necessary to sustain those relations and how do people incorporate them into their everyday lives? As we illustrate with the case study from Bangladesh below, concerns over algorithmic fairness can be operative even where there is no operational algorithm. Indeed, our research subjects intuitively understood their experience as a matter of algorithmic fairness—does this system compare them to the local and international collective of tests takers in a fair and understandable manner?—despite the distinct lack of a sophisticated machine learning component. A potential way through this paradox is a situated fairness approach guided by the interests of the impacted stakeholders.

## 2.3 Understanding Algorithmic Systems through Perceptions and Folk Theories

Questions of autonomy, power, and agency of data subjects—people who are "both resources and targets" for algorithmic systems [111:2]—are emerging as topics of research in FAccT and allied fields [22, 52, 54, 72]. Studies have suggested new ways for data subjects to 'know' algorithms through improved measures for accountability and fairness, including disclosure requirements, regulatory oversight, and explainability [28, 57, 65, 89, 97]. This research has expanded to consider the role that data subjects themselves might play in generating understanding of the workings of algorithmic systems. While ordinary people may not be able to account for algorithms in terms of statistics and code, they still develop an idea of how these systems work and impact their lives. Studies of data subjects' folk theories of algorithms have tended to focus on mundane sense making around how social media algorithms curate their newsfeeds [15, 18, 34]. Central to these studies is unpacking concerns that trigger such sensemaking and engagement with algorithmic systems that can range from publicly discussing their experience [such as in 15, 18, to "everyday audits" and pursuing remediation [95].

A key trigger for such engagement is the perceptions of data subjects around being treated unfairly by algorithmic systems. Research focusing on their perceptions often relies on building speculative cases for appropriation of algorithmic systems to encourage data subjects to articulate their understanding of fairness and thereby create boundaries around what an algorithmic system should or should not do [12, 43, 61, 109, 110]. Our research draws on and contributes to this literature by exploring a unique situation in which an algorithm recedes from the collective imagination of a data subject community because it is no longer deployed. Prompting participants to discuss the impacts of an algorithm brought about conscious reflection on the extent to which the algorithmic processes—which most thought had been eliminated when the algorithm was no longer in use—continued to structure their lives.

## 3 METHODS

The following chronology of events has been weaved together through a series of 47 interviews taken since July 2021 with 33 students and 14 teachers from Bangladesh. Beyond the formal interviews, since September 2020, we had 103 informal conversations with stakeholders. The preliminary informal conversations established rapport and trust with community members, many of whom were initially hesitant to share out of fear of retribution. We also engaged with community members through online groups, amplifying their voices whenever possible. Through a steady process of community engagement through local contacts, we gradually succeeded in gaining the trust of our participants.

Most of our participants are based in Dhaka (the capital of Bangladesh) but some also come from other cities like Chottogram (Chittagong), Narayanganj, and Sylhet. Amongst the students, 19 are Cambridge International Examinations (CIE) candidates and 8 are EdExcel candidates (further details in Sec. 4). All teachers had at least five years of experience teaching experience for GCE O and A Level. All informants are 18 years or older.

After providing informed consent to participate, each interviewee took part in semi-structured interviews that lasted 62 minutes on average. 44 of 47 participants engaged in two rounds of interviews, first at the start of the project and second towards the end. During the first round of interviews, participants provided (a) perspectives on the exams (shared in Sec. 4) and (b) information around the events surrounding the algorithmic grading, which forms the timeline presented below. In the second round, participants were encouraged to critically reflect on the events again while engaging with our prompts that brought the algorithmic imprint into focus. To protect the privacy of our informants, we have redacted identifiable details and added pseudonyms where applicable. The interviews were primarily in Bangla, punctuated by sessions in English or Banglish (a hybrid of Bangla and English). Two Bangla-speaking researchers did the translation. Many of our participants shared email correspondence to corroborate their verbal accounts of the sequence of events. Two researchers used an open-coding scheme to iteratively conduct Thematic Analysis [6, 14] on the data, culminating in themes presented in Section 5.5 around critical engagement with the imprint.

We focus on GCE A Level for the June 2020 session because it portrays a rare opportunity to analyze the state of the world with and without algorithmic grading. For the sake of tractability and reducing conflation, we will use CIE's timeline as our baseline for two reasons: first, teachers who taught both CIE and EdExcel students felt that CIE did a better job with regular updates. Second, EdExcel's timeline is roughly the same and lags the CIE timeline.

## 4 CONTEXT: GCE EXAMS, & BANGLADESHI PERSPECTIVES

There are two milestone exams—GCE O(ordinary) Level and A(advanced) Level, typically taken in grades 10 and 12, respectively. In addition to International Baccalaureate (IB), GCE exams are the predominant international high school certificates accepted by universities worldwide [115]. In Bangladesh and beyond, there are two UK-based boards that administer these exams—Pearson EdExcel (or EdExcel) and Cambridge International Examinations (CIE). Both boards are subject to the regulatory authority of Ofqual, a non-ministerial government department in the UK [117]. For students in the GCE curricula, the O and A level exams are two of the most consequential exams of their lives. "GCE A levels are the make it or break it exam. You can be a bad student till your 11th grade and if you shine in A levels, the past doesn't matter" (S19). The A level examination is particularly important since the results govern not just university placements within one's country but also opportunities for both local and international scholarships [9, 75]:

> This might be vulgar, but it feels like I'm paying money to my colonizer in the UK for a piece of certificate that tells the world I am no dumber than a local UK kid or that I'm competitive at a US university. Sometimes, it's hard to ignore that reality. If our local [Bangladeshi] curricula were well-accepted worldwide, my parents might not have enrolled me for the GCE system (S21, emphasis added).

There are two main institutional nodes in GCE exam administration outside the UK. First is the *British Council*, the UK's international

">The Algorithmic Imprint                                                                                                       FAccT '22, June 21–24, 2022, Seoul, Republic of Koreaorganization for educational opportunities, which are located all over the world; the one in Bangladesh is 70 years old (the country itself is 50 years old as of 2021). The British Councils interface with the UK boards and administer the tests at the country and district levels [118]. The second node is comprised of the *schools or centers* through which students register for these exams. There is a variance in the number of subjects a student takes; for GCE O Level, our participants took between six to fifteen subjects and for A-Level, they took two to six subjects. Dhaka, the capital, has 115 out of the 146 schools (centers) in Bangladesh [32, 112].

The *registration costs* vary between boards. For each subject, participants reported registrations costs of approximately BDT 48,800 (USD 574) [119]. The costs can add up quickly; a student taking exams for four subjects must pay ~$2300 to take A Level exams. Most "students sitting for GCE exams are from the lower middle to the upper classes in society. For the rich, it's nothing. For everyone else, it might require saving for years or taking a loan to not just pay for registration but also the associated school and coaching center [cram school] fees" (S25). In comparison, SAT registration fees are around $65 while AP is $40 per subject.

*Practices of studying* for these exams also *differ from region to region.* In Bangladesh, students "typically take it easy in the school during the year and ramp up studying in the last few months" (S6). Given the exam results are what counts as measure of academic competence, "not every student takes the schoolwork as seriously as students in other curricula might take. If you do well in the exams, all 'sins' are forgiven" (S14). As a result, there is a level of "cramming in the last 30-60 days before the exam begins. Some of [the students] also strategically use the naturally occurring breaks between subjects [to cram] during the exam period, which can last for up to 1.5 months for A Levels" (S3).

## 5 SITUATING THE ALGORITHMIC IMPRINT: CHRONICLING THE EVENTS IN BANGLADESH

We situate the algorithmic imprint by chronicling events and a timeline of events between March 15 to Aug 17, 2020. This period covers the start of the Covid-19 lockdown to the end of the revised grading process in Bangladesh. To facilitate their traceability and understandability of the interconnected events, we have divided the timeline into four acts that correspond with most participants' accounts. Each act marks significant events that shifted the conditions and practices involved in the algorithmic grading of A level exams:

- The first act covers the period between March 15 to March 30, 2020, when the decision to eventually cancel the exams during a pandemic-induced lockdowns.
- The second act, spanning April 1 to August 10, 2020, focuses on efforts around alternative assessments mechanisms (e.g., historical grades, rank order, and algorithmic standardization) and their limitations.
- Focusing on a single day, August 11, 2020, the third act recounts the protests and chaos on the "Results Day" when the algorithmic grades were released.
- The fourth act, between August 12 to August 17, focuses on the aftermath of the protests and removal of the algorithmic

grading with a critical reflection on whose voice mattered in enacting these revisions.

We illustrate imprint of algorithmic grading by highlighting the conditions and practices that made its deployment possible and how these conditions and practices persisted even when it was removed. The events we describe co-constitute [51, 67, 70] in the timeline of these events. To illustrate the imprint and engage with it, we rely on two main sources: (1) the accounts of our research participants (Sec. 5.1 to 5.4 chronicling the timeline), and (2) their self-reflections when prompted to think through the conditions and practices of the algorithmic mediation (in Sec. 5.5). We conclude each act (Sec. 5.1-5.4) by the implications of these acts for the imprint.

### 5.1 First Act: Lockdown and the flipflop around exam cancellations (March 15- March 30)

The first act describes the *pre-conditions* needed for the algorithm to come into existence and its resulting imprint. Around March 15, 2020, Bangladesh went into lockdown due to Covid-19 global pandemic and chaos ensued around school closures. Overnight, students faced a "chaotic and anxiety inducing" (S22) situation, particularly around how they would study at home and how the exam would be administered in three months' time. The month of March is a "ramp up month when most of the mock exams take place" (S4). The lockdown left students "in a limbo" (S8) while teachers "faced intense levels of anxiety not knowing how to prepare [their] students" (T8). Compounding these challenges, "online learning infrastructure is basically non-existent... and data is expensive [in Bangladesh], making the pivot to remote learning an extremely difficult task. [Teachers] had no idea how use all these tools to teach remotely... Learning is always in person in Bangladesh, not like the US where you've a history of online classes" (T3).

Exacerbating the uncertainties was the flip flop around cancellations of exams. On March 20, CIE cancelled exams in the UK but not internationally. Bangladeshi students felt alienated: "CIE only cared about the safety of the UK students. We live in a third world country. Why bother with our safety, right? All they care about is our money" (S9). However, three days later around March 23, CIE reversed its initial decision and cancelled exams globally. The news of exam cancellation was met with mixed feelings. Students and teachers were simultaneously relieved and stressed out. On the one hand, students "felt relieved because [they] wouldn't have to risk exposing [themselves] or [their] families to the coronavirus" (S13). On the other hand, they were also anxious; "we didn't know what's going to happen—would we lose a year of progress? How would they grade us if we don't have exams?" (S5). From teachers' perspective, they too appreciated being able to protect their families but worried about assessment.

In summary, the pandemic-induced lockdowns produced the *pre-conditions for an algorithmic intervention and by extension, its imprint,* triggering a series of interventions in existing educational practices in Bangladesh. Awareness of pre-conditions can help us better understand how the contours of the imprint shape stakeholders' lives.

## 5.2 Second Act: The Long Chaotic Journey of Alternative Assessments (April 1 – August 10)

In this act, we describe the data and organizational practices imposed by the algorithmic intervention and their roles in the imprint. Around April 1, 2020, teachers learned about 'alternative assessment" methods (T1, T12). *Students would be evaluated using three elements*: (1) their past performance to generate an estimated grade for a subject (called a *Teacher Assessed Grade (TAG)*), (2) their rank order for each subject, and (3) an algorithmic standardization that uses the personal data (rank order and TAG) but also historical data at a center (school) level. As we will see, even though only the third element can conventionally be considered "the algorithm", TAG and rank order are crucial for it to work and thus are implicated in describing its imprint. Although these measures were touted as "being fair and objective measures by the boards" (T13),they are grounded in *two problematic assumptions*: (1) past performance is predictive of future competence; and (2) no two students can be equally competent in a subject (no ties allowed in ranking). These two assumptions are crucial to implementing algorithmic grading and making sense of its imprint.

*5.2.1 Assessment Element 1: Teacher Assessed Grades (TAGs) & Historical Information.* According to the exam boards, the teachers could use sources like "mock exam results, past classwork performance, previous board exam grades" (T6). While this strategy sounds reasonable in theory, but it is also problematic:

> This is where the main disconnect happens. The UK folks have no idea of the culture of learning [in Bangladesh]. Students barely start taking this seriously till the last 60 days when the real preparation begins. So, anything before that [time] is not representative, is it? Also, how many schools actually maintain these records rigorously? We aren't like the UK where everything is digital. [CIE] wants the data. Fine. But where will we get it? Make it out of thin air? This whole historical data thing is a mess (T5, emphasis added).

The teacher's insights expose the mismatch in existing educational infrastructural conditions in the UK and in Bangladesh. They also highlight the differences in situated practices of learning between the two ecosystems. *Bangladeshi students often take a different path to the exam*, "contrary to what CIE assumes [the] students do" (T6). Such a culture of exam preparation is not uncommon in South Asia, where "exams are the end all be all... Learning culture-wise, day-to-day classwork doesn't really matter. Things are different from 'Western' countries"' (T13). In other words, unlike a "build-up culture" (S8) where daily course materials add up throughout a term, the exam preparation culture in Bangladesh "happens in bursts with everything culminating in the last 30-60 days" (S17). Given this understanding, we can see how constructing a historically-normed grade is problematic. Since no one can "change the past, the historical grades made [the students] feel as though [the exam boards] *changed the rules of the game after the game started"* (S14, emphasis added). Many students felt "helpless and voiceless" (S2, S9, S25). They felt they were "robbed of the agency to build [their] own future" (S11) in the absence of the exam—"the great equalizer" (S8). Furthermore, only a minority of schools affiliated with our participants kept "rigorous records of students' past performance [because] it never mattered in the past" (T9). When asked if the exam boards sought their input on how to formulate alternative assessment, everyone responded with a resounding, "No!" Teachers and students alike were frustrated by the imposition of these assumptions and contextualized their experience through the lens of colonialism:

> Not everyone takes the same path to reach the same destination. [Bangladeshi] students do just as well if not better than those in the UK. We know our methods aren't bad, but does the 'mothership' in the UK care?! To them, we are just another third world country. We must do what the colonial masters ask (T7)

The requirement produced a rush to assess students to *compensate for data voids* on *historical grades*. The rush, in turn, created conditions for emergent mechanisms of arbitrary assessment, putting disproportionate burdens on students. To "make up for the lack of historical data" (S13), many centers had to impute the data. "Every center ended up doing its own thing" (T9), which created confusion amongst students. With "vague instructions that clearly weren't made with Bangladesh in mind" (T3), students were presented with arbitrary assessment material. For example, many were asked to take online mock exams in which one would be "solving the past question papers from the last 5-10 years. If you have 5 subjects, that's 25-50 years' worth of question papers" (S21). Many opted to use mark schemes (answer keys) to solve these papers. While that would technically count as cheating, students felt that in "these unfair circumstances, that is the fair thing to do" (S16). Their reaction had an element of subversion aimed at regaining their voice and power: "CIE was trying to f**k us up, so why can't we f**k them back?" (S24). Notably, students did not blame their teachers. As one student put it: "Teachers are powerless here. The UK boards gave them no support" (S8). Teachers acknowledged the "imperfect nature" (T10) of these assessments as well.

*5.2.2 Assessment Element 2: Rank order of students.* Beyond the TAGs, teachers were also asked to provide a *rank order* of students, "and no ties were allowed no matter how similar the students were" (T8). Neither the students nor the teachers knew *why the ranking had to be done in an ordinal manner or how it would be used.* As we will eventually witness in Sec. 5.4 and 5.5, the *rank order will play a crucial role in the algorithmic imprint* (Sec. 5.4) and *become a focal point of critical reflections of participants* (Sec. 5.5). With the enforcement of ordinal ranking, the teachers "felt trapped. [They] had to obey the orders from the UK boards but [they] struggled with the ethics of ranking students like this. It didn't feel fair" (T5). A traditional exam allows for the possibility of ties in scores. With the *forced monotonic ordinal ranking*, the statistical manifestation of "fairness" is changed by a unilateral decision from the exam boards, creating a new hierarchy between students through exam scores. Every teacher corroborated this ordinal ranking requirement. Some shared emails corroborating this instruction.

Between the first week of April and mid-June 2020 (when grades were due), students and teachers juggled through multiple avenues of assessment. The boards built a global platform for teachers to submit their grades. Teachers had to download a "preformatted CSV file that caused more trouble than it solved" (T6). Confronted with a "confusing series of instructions" (T8), teachers ultimately submitted their TAGs and rank orders by the June 16 deadline. Students also suffered from "miscommunication during the April to June period because the boards themselves sent conflicting information around grades" (S27). Moreover, teachers felt that by putting the responsibility on them to rank students, the boards were protecting themselves and making teachers share "culpability if something went wrong" (T14). They were made into potential "scapegoats; students could reach out to teachers instead of the faceless exam boards" (T2).

*5.2.3 Assessment Element 3: Algorithmic Standardization.* The TAGs and ordinal ranking were created in service of the final facet of the alternative assessment—the *algorithmic standardization*. The following four factors situate the complexities around this standardization:"

First, *no procedural details around the standardization were shared before the grades were due* in mid-June 2020. Around July 31, the boards shared the *first communication* around the algorithmic procedures. By this time, most people "were exhausted. No one really cared what the hell they'd do" (S26). Notably, *the term "algorithm" was never used in any of the communication*. Participants perceived the omission of the word "algorithm" as a preemptive "tactic to deter criticism" (S3). Moreover, only two out 47 informants recognized the algorithmic equation during our conversations. Given how well-informed our participants were, this was surprising. Participants perceived the lack of disclosure of the algorithm as intentional, insisting that the boards "wanted to hide everything till we were exhausted and tired with the process. Releasing details after the grade deadline also meant there was nothing [they] could do about it" (T2).

Second, there was a controversial move to *use of past performance data at the cohort/center level.* Beyond the individual student data, the standardization process would take "the performance of previous cohorts from the same school into account" (S20). Hence, a student's grade could be severely impacted if previous cohorts performed poorly, or if the school lacked historical data. While the intent behind this might have been to deter schools from providing unrealistically positive grades, the implications of potentially "punishing high performing cohorts with previously low performing ones felt unfair" (S12). In the absence "school level historical data, the boards would use regional or global averages, which would be disastrous for [Bangladeshi] students" (T7), especially considering the differences in educational infrastructures.

Third, *many teachers made an (unfounded) assumption around how the algorithm would work*—they expected that the algorithm would "intelligently bump [their] TAGs because [they] always gave lower marks on mock exams to motivate the student to study harder" (T4). Note that this is a "well-established cultural practice and students expect the harsh grading" (T12). This assumption around 'bumping-up' grades was the most prevalent folk theory. Given the boards did not divulge algorithmic details till grades were due, it was difficult for the teachers to be informed about how the algorithm worked. Teachers overestimated the algorithm's abilities: "It's clear that the algorithm was stupid. Everyone knows we mark the mocks strictly. We thought the system was intelligent" (T3).

Fourth, and most importantly, no one was aware that the *major reason behind demands for the non-tie-based ordinal rank order was to accommodate data inputs to the algorithm.* It was not that the exam boards did not believe in ties. Rather, the algorithm, to function properly, needed ordinal rank data with no ties as input. The standardization process imposed data practices on teachers to make student performance commensurate with the algorithm's inputs. The algorithm imposed a new set of infrastructural requirements from the educational ecosystem of Bangladesh: ordinal ranking, historical grades to generate TAGs, and school-level historical data. When we brought these infrastructural requirements to their attention, our participants expressed a mixture of disbelief and frustration:

> Holy shit! I always wondered why they pushed us to do [the ranking] in this manner... This also makes me so angry. To force us to do all these things just so that an algorithm could work is insane. No wonder they hid the algorithmic details from us. (S4, emphasize added)

In a nutshell, this act illustrates *how algorithms living in the digital world often shape data and organizational practices in the real-world that persist well beyond the algorithm's deployment.* The TAGs and ordinal rank order are rooted in typical algorithmic interventions: (1) using the past to predict the future, and (2) using group data to make judgements about individuals. They are also emerging infrastructural practices that made standardization possible. *The practices centered on producing them are the core aspects of the algorithmic imprint* that continued to have a lasting impact on the stakeholders, persisting even when the standardization was removed (Sec. 5.4 and 5.5). Stakeholders' folk theories of algorithmic performance created major downstream effects (over-estimating the algorithm's abilities) from upstream requirements (ordinal ranking).

## 5.3 Third Act: Results, Protests, and Villainization (August 11)

On Aug 11, the *algorithmically calculated* A level results (hereafter referred to as Round 1) were announced. The announcement was met with worldwide protests. On the one hand, risking arrests, Bangladeshi students "protested because [they] felt they had been robbed off their future" (S26). This student shared the psychological impact of the announcement:

> I felt so helpless. What'd I tell my parents? They invested so much money, and this is what I've to show for? They won't even understand all the algorithmic stuff. I knew I had to go to protest. This was not fair (S27).

The teachers, on the other hand, faced backlash from students. Students could not reach out to "faceless organizations that were pulling strings in the UK" (T9). The teachers, being the only accessible party, felt like a "punching bag, taking the blow for something [they] had no voice in" (T4). As another teacher recounted:

> I've never felt like a villain in my 20+ years of teaching. My phone had 83 missed calls in 2 hours that day. My own students, my own kids, were blaming me. Problem is, I had no agency—I couldn't even reach out to the

> boards. It was pure chaos. It made me question my self-worth as a teacher. I've never ever done that. (T12).

The third act highlights how certain lived experiences can shape stakeholders' sensemaking of algorithmic operations. These lived experiences have deep and persistent psychological impacts. Participants had vivid recollections of the Results Day and how it shaped their experience of algorithmic grading. Understanding the imprint requires that we pay attention to these events and identify the practices that matter most to data subjects in their lived experiences of algorithmic systems. After all, *these conditions and practices shape the contours of the imprint in the first place.*

### 5.4 Fourth Act: Revised Grades and the Algorithm's Afterlife (August 12 onwards)

The final act illustrates one of the most important aspects of the imprint: *how it persists even when the algorithm is removed.* On August 17, 2020, CIE announced that they are *rescinding all grades globally*, "a once in a lifetime event" (T8). They shared that "no one will receive a grade lower than the TAG" (S27). *This meant that the algorithmic standardization was no longer in use,* although the boards never "explicitly acknowledge they were removing the algorithm. It's as if they knew they screwed up but lacked the courage to own up to their mistakes" (S19).

When the grades were revised (hereafter referred to as Round 2), they "improved across the board" (T10). It was clear that TAGs were more "in line with student expectations than the algorithmic adjustments" (S17). In Bangladesh, many students actively apologized to teachers. The teacher who felt villainized (in the previous subsection) reported:

> The flowers and apologies meant the world to me. I chose to be a teacher because I love helping students. I am still upset, but my anger is towards the boards who used us as human shields. (T12, emphasis added)

Despite all the struggles, this story, on the surface, appears to have a happy ending—the algorithmic standardization was removed and the "majority of the students were happy with the grades" (T10). The grades of all 33 students increased or remained the same. However, *was the algorithmic intervention truly undone?* Recall that students were never re-graded from scratch. Every teacher underscored how "the need to rank students without ties fundamentally impacted how [they] assigned certain grades to students" (T9). The rank order was a vestige of the algorithmic processes that fundamentally influenced TAGs. Without regrading from scratch, *both these elements persisted as imprints of algorithmic grading even when the standardization step was removed.* Teachers highlighted how it appeared "as if *the ghost of the algorithm* remained when the revision happened" (T12). We prompted teachers and asked: if you did not have to rank your students as the algorithm required, would you have graded them differently? Every teacher answered with a resounding, "Yes"! One of the teachers aptly summarized:

> We changed the way we graded because the standardization required us to. I'd never rank my students like this. When the grade revisions happened, we didn't assign new TAGs. Students still had the old TAGs, which meant the ranking is still there. So, *the ghost of the algorithm is still there* (T5, emphasis added).

*Like the remnants in a palimpsest, the imprint of an algorithm persists even in its afterlife.* The frequent recurrence of "the ghost of the algorithm" in our conversations profoundly shaped our conception of the algorithmic imprint. At the surface, it might seem that, as the exam boards claimed, simply removing the algorithmic standardization resolved the problems of socio-statistical bias in received grades. However, further reflection reveals a different picture: while the standardization was removed, *the algorithmic processes were never undone.* There is no simple "undo" button for algorithmic deployments.

### 5.5 Critical Engagement with the Algorithmic Imprint

We switch gears and critically engage with the concept of the algorithmic imprint by presenting the emergent themes from the *second* round of interviews. During the second round of interviews, to encourage critical reflection among participants on their struggles, we focused more explicitly on using the concept of the algorithmic imprint and related insights that emerged from the first round. This self-reflective exercise generated thematic insights on *key affordances of the algorithmic imprint* for different stakeholders to make sense of algorithmic interventions. Below we share three thematic aspects of imprint-aware participants' critical reflections: (1) the frequent *invisibility of people as infrastructure* [60, 98] and *how the imprint can be a resource to account for invisible labor*; (2) struggles with a *lack of agency* and voice and *how the imprint can act as an anchor for enacting accountability*; and (3) critically *questioning the very need for algorithms* and *how the imprint can be a guiding principle to promote mindful deployments.*

First, starting with *invisibility of people as infrastructure*, teachers shared that they received no recognition or compensation. The boards did not "pay a single Taka [Bangladesh's currency] despite making [the teachers] to do the work" (T2). The hidden labor was unnecessarily harder because the algorithm required ordinal ranking with no ties. With the eventual removal of standardization, their work became even more invisible. Seeing their efforts through the lenses of imprint brought these invisible elements back into to conscious reflection, creating affordances for the *participants to account for their invisible labor*. As one teacher put it: "this imprint thing you described helps me in two ways—it helps me understand how my efforts were exploited by the boards. It also makes me self-aware of my own contributions" (T8). Without the notion of the imprints, teachers felt "their hidden labor would've been impossible to identify" (T2).

Second, participants often shared their *struggles with lack of agency and voice* as the process of algorithmic grading unfolded. Although the revised grades were mostly better than algorithmic grades, most reported that they were left with a "bad aftertaste" (S18) and a *deep sense of injustice.* While they "appreciated the improved grades, [they] could not ignore what the process had put them through" (S19). When prompted to reflect on their experience using the vocabulary of imprint, students, just like their teachers, noted that it helped them pinpoint "the source of the sadness [they] felt even if the grades improved" (S3). Reflecting on the concept, this student expressed a reconciliatory moment:

> *Despite my grades improving, something always felt off. The metaphor [of algorithmic imprint] gives me a thread to connect the dots now. The imprint helps me track the pain and suffering. It can never be erased. It helps because now I know who to hold accountable—it isn't the teachers. It's the UK boards.* (S17)

The imprint, thus, offers a way "to understand the hidden sources of injustice even when things look good on the surface" (S14). It can act as an anchor for enacting accountability and situate struggles of living with algorithmic systems, affording traceability to capture effects otherwise hard to detect. It can facilitate a deeper understanding of how the very infrastructural conditions and practices that make such systems possible can also be unfair in their own unique ways.

Third, the imprint helped our participants critically *question the very need for algorithmic deployments*, a sentiment expressed by this participant: "After all the hassle, we now know that we could have done the whole damn thing without the algorithm. So *why did we add it in the first place*" (S14, emphasis added)? Even though the algorithmic standardization was removed, the participants could not ignore how "an unfair infrastructure was set up [for it] burdening people" (T5). Teachers felt that imprint-awareness "would have made the bosses in UK think twice before going ahead with this mindless catastrophe" (T14), highlighting how imprint can serve as an ethical principle for mindful deployment and appropriation of algorithmic systems.

## 6 IMPLICATIONS OF THE ALGORITHMIC IMPRINT

The notion of the algorithmic imprint has practical and conceptual implications. In this section, we highlight three implications around how imprint-awareness can (a) reframe how we think about algorithmic impact, (b) inform how we design algorithms, and (c) guide us in AI governance. These implications are not intended to be exhaustive; rather, a starting point for a critically constructive discourse on algorithmic deployments.

*First,* being aware of the imprint-laden nature of algorithms can *broaden how we can assess algorithmic impacts*. Typically, FATE-related impacts are most salient and addressed during the lifetime (or use) of the algorithm [86]. As our findings illustrate, just because an algorithm is removed, it does not mean its effects are undone. The concept of the imprint can empower stakeholders to address these issues in the afterlife even when the algorithm is removed. The death of the algorithm does not entail the death of the issues it created. Moreover, algorithmic impacts can be harder to detect in the algorithm's afterlife because its absence can make these issues invisible. The fact that there is no easy "undo" button for algorithmic deployments implies that, on the one hand, developers and operators need situated fairness perspectives around the tangibility of infrastructural conditions and practices that emerge to sustain algorithmic interventions. On the other hand, activists and regulators must also acknowledge the persistence of these practices in approaches to remediate algorithmic harms and demand accountability beyond removing algorithmic components of a system. The notion of the imprint provides a tractable pathway to include the socio-political factors around the technical algorithmic system in our conceptualization of algorithmic harms and biases. It extends the locus of analysis and becomes a junction between use (lifecycle) and post-use (afterlife), broadening the assessments of algorithmic impact.

*Second,* attention to hard and persistent infrastructural impacts can facilitate *imprint-aware algorithm design*. Designers and developers can utilize the imprint as a guiding tool to think about the footprint of their creations and design to mitigate harm. Software, by its nature, is malleable, reversible, and mutable [19, 49]. Lines of code are easily commented out, deleted, and their digital effects reversed easily [55, 87]. It may be illusive to think that because algorithms are made of software, their effects are transient. However, as our findings highlight, it is anything but that. These digital lines of codes in algorithms can leave imprints in the data infrastructure in the real-world and leave their mark on stakeholders at the cognitive, behavioral, and emotional levels. As we saw in our case study, an algorithm that benignly requires ordinal scores can permanently and harmfully alter the relationship between teachers and students in a well-established educational culture. What if the developers of the Ofqual algorithm were aware of the potential impact of their algorithms outside the bounds of algorithmic grade standardization? If algorithmic designers are aware that their digital creations can leave persistent imprints in the real-world, they can configure the infrastructure such that such that upstream changes (e.g., parameter tuning, data inputs) [29, 31] in the algorithm's makeup do not cause harmful downstream effects on stakeholders (e.g., unfair treatment) [30]. Even if there are drastic changes, the imprints can be designed to be made *more explicit* so that people subjected by them are more aware of them. Awareness and traceability of the contours of the imprint can facilitate improved active participation in algorithmic mediation [62]. *An imprint-aware design mindset treats stakeholders as active (as opposed to passive) participants in the design process.* If designers are imprint-aware and do not assume algorithmic impacts are easily undone, they may design things differently. For instance, developers can add checkpoints to the development cycle that explicitly tackle different scenarios arising from divergent forms of imprints. Teams can take a participatory approach with stakeholders and utilize techniques from HCI such as scenario-based design [88], Reflective Design [30, 94], and Value Sensitive Design [37] to mitigate the ill-effects of imprints. Lessons from the algorithmic afterlife can reflexively inform an imprint-aware design mindset of algorithms resulting in better accountability in deployment. The imprint-aware design mindset, in turn, can make the algorithmic development process more human-centered and sociotechnically-informed.

*Third,* algorithmic imprints have sociotechnical complexities that require *interventions at the governance level*. Technical interventions alone cannot mitigate harms from algorithms. Thus, we need to complement imprint-aware algorithm design with appropriate AI governance [23]. Organizationally, imprint-awareness can be incorporated into existing AI ethics literacy programs [66] similar to recent work demonstrating that literacy of dark patterns can promote self-reflection and mitigate harms [68]. Algorithmic imprint literacy programs can empower (a) designers and developers to proactively mitigate harm and (b) end-users to be proactively aware of them. These programs can include simulation exercises using speculative design [8] and reflective design [94] to envision "what could go wrong" [20] from the technical and infrastructural perspectives. Many algorithmic governance mechanisms have been developed for the machine learning lifecycle, but these mechanisms are largely focused on the technical components most proximate to the activities of data scientists. None yet facilitate governance or consideration of data infrastructures that precede the construction of datasets, which are increasingly recognized as ethically consequential [48, 96]. Moreover, an imprint-aware mindset can proactively *guide legislation and regulations* around algorithmic deployments. As we step into a future where more algorithmic deployments are stopped, we need the necessary regulatory foresight to address imprints in the algorithms' afterlife. For instance, recently Facebook has stopped using its facial recognition algorithm [114, 120]. However, its imprint remains in the models trained with the data and manifests in multitudes across other Facebook-owned platforms where the algorithm is still in use [46]. As regulators increasingly turn toward transparency mechanisms such as algorithmic impact assessments [2, 76], there should be some incorporation of infrastructural perspectives into such assessments. The notion of the imprint provides practical tools to broaden our conception of algorithmic impact, especially in the algorithm's preceding infrastructures and in its afterlife, as effects linger.

## 7 CONCLUSIONS

In this paper, we introduce the concept of the *algorithmic imprint* to illustrate how algorithmic consequences and harms can extend well beyond the algorithm's lifetime. The concept of the imprint provides the vocabulary to talk about the algorithm's afterlife, an uncharted domain. We practically situate the concept of the imprint through the events in Bangladesh around the Ofqual algorithmic grading of the June 2020 GCE A-Level exams during the Covid-19 pandemic restrictions. Chronicling more a year-long community engagement consisting of 47 interviews with teachers and students, we present the first coherent timeline of not only what happened in Bangladesh but also contextualize the events through the lenses of the imprint. Integrating student and teacher perspectives, we provide interweaving layers of context, adding novel perspectives about the social and ethical consequences of the data infrastructures and social structures necessary to support algorithmic grading. In analyzing the case study, we deploy the notions of an *infrastructural lens* (foregrounding the infrastructures necessary to support algorithmic systems) and *situated fairness* (the prioritization of the lived experiences of impacted stakeholders in the construction of algorithmic fairness measures). Despite the removal of the algorithmic standardization of the exam grades, we illustrate how and why the algorithm left persistent imprints at the infrastructural, social, and individual levels. Like the remnant traces on a palimpsest, the imprints of the algorithm can persist long after its deployment period. Recognition of the imprint-laden nature of algorithms provides a means to theorize how the impacts of an algorithmic system exceed the technical boundaries and the lifetime of that system. Thus, it broadens how we assess algorithmic impact. Imprint-awareness can allow us to identify implications that inform algorithm design for designers and developers and guide legislation and regulations for policymakers engaged in AI governance. Thus, an imprint-aware mindset can make algorithmic deployments more human-centered and sociotechnically-informed. Not paying attention to algorithmic imprints can inhibit our abilities to sufficiently tackle a future where more algorithmic operations are discontinued yet their harmful effects persist.

## ACKNOWLEDGMENTS

With our deepest gratitude, we acknowledge the time our participants generously invested in this project. We are grateful to them for courageously sharing their lived experiences and adding the Bangladeshi perspective to the narrative around the Ofqual algorithmic grading controversy. We sincerely thank Abrar M Fuad for his invaluable contributions in facilitating community connections in Bangladesh and facilitating the field work. We are grateful to members of the Data & Society Research Institute and participants of the Parables of AI In/From the Global South whose input refined the conceptualizations presented here. We are indebted to Soledad Magnone, Michael Muller, Karen Hao, Divy Thakkar, Samir Passi, and Jenna Burrell for their generous and detailed feedback at different stages of the project. Special thanks to Naima Rashid, Intekhab Hossain, Azmain Amin, Dewan Lamisa, and Jeni Tennison for their engagement and constructive feedback during the early stages of the project. This project was partially supported by the National Science Foundation under Grant No. 1928586.

## REFERENCES

[1] Rediet Abebe, Solon Barocas, Jon Kleinberg, Karen Levy, Manish Raghavan, and David G. Robinson. 2020. Roles for computing in social change. In *Proceedings of the 2020 Conference on Fairness, Accountability, and Transparency* (FAT* '20), 252–260. https://doi.org/10.1145/3351095.3372871
[2] Ada Lovelace Institute. 2020. *Inspecting Algorithms in Social Media Platforms.* Retrieved from https://www.adalovelaceinstitute.org/algorithms-in-social-media-realistic-routes-to-regulatory-inspection/
[3] Bobby Allyn. 2020. IBM Abandons Facial Recognition Products, Condemns Racially Biased Surveillance. *NPR.* Retrieved January 4, 2022 from https://www.npr.org/2020/06/09/873298837/ibm-abandons-facial-recognition-products-condemns-racially-biased-surveillance
[4] Louise Amoore. 2020. Why "Ditch the algorithm" is the future of political protest. *The Guardian.* Retrieved January 14, 2022 from https://www.theguardian.com/commentisfree/2020/aug/19/ditch-the-algorithm-generation-students-a-levels-politics
[5] Sareeta Amrute. 2020. Bored techies being casually racist: race as algorithm. *Science, Technology, & Human Values* 45, 5: 903–933.
[6] J Aronson. 1994. A pragmatic view of thematic analysis: the qualitative report, 2,(1) Spring.
[7] Chinmayi Arun. 2020. AI and the Global South: Designing for Other Worlds. In *The Oxford Handbook of Ethics of AI*, Markus D. Dubber, Frank Pasquale and Sunit Das (eds.). Oxford University Press, New York, 589–606.
[8] James Auger. 2013. Speculative design: crafting the speculation. *Digital Creativity* 24, 1: 11–35.
[9] Douglas G. Bagg. 1968. The Correlation of Gce a-Level Grades with University Examinations in Chemical Engineering. *British Journal of Educational Psychology* 38, 2: 194–197. https://doi.org/10.1111/j.2044-8279.1968.tb02005.x
[10] Solon Barocas, Moritz Hardt, and Arvind Narayanan. 2019. Fairness and Machine Learning: Limitations and Opportunities. Retrieved from http://www.fairmlbook.org
[11] Emily M. Bender, Timnit Gebru, Angelina McMillan-Major, and Shmargaret Shmitchell. 2021. On the Dangers of Stochastic Parrots: Can Language Models Be Too Big?. In *Proceedings of the 2021 ACM Conference on Fairness, Accountability, and Transparency*, 610–623.
[12] Reuben Binns, Max Van Kleek, Michael Veale, Ulrik Lyngs, Jun Zhao, and Nigel Shadbolt. 2018. "It's Reducing a Human Being to a Percentage": Perceptions of Justice in Algorithmic Decisions. In *Proceedings of the 2018 CHI Conference on Human Factors in Computing Systems.* Association for Computing Machinery, New York, NY, USA, 1–14. Retrieved December 7, 2021 from https://doi.org/10.1145/3173574.3173951

[13] Abeba Birhane. 2021. Algorithmic injustice: a relational ethics approach. *Patterns* 2, 2: 100205. https://doi.org/10.1016/j.patter.2021.100205
[14] Virginia Braun and Victoria Clarke. 2006. Using thematic analysis in psychology. *Qualitative research in psychology* 3, 2: 77–101.
[15] Taina Bucher. 2017. The algorithmic imaginary: exploring the ordinary affects of Facebook algorithms. *Information, Communication & Society* 20, 1: 30–44. https://doi.org/10.1080/1369118X.2016.1154086
[16] Joy Buolamwini. 2019. Response: Racial and Gender bias in Amazon Rekognition — Commercial AI System for Analyzing Faces. *Medium*. Retrieved October 5, 2020 from https://medium.com/@Joy.Buolamwini/response-racial-and-gender-bias-in-amazon-rekognition-commercial-ai-system-for-analyzing-faces-a289222eeced
[17] Joy Buolamwini and Timnit Gebru. 2018. Gender Shades: Intersectional Accuracy Disparities in Commercial Gender Classification. In *Proceedings of Machine Learning Research*, 1–15. Retrieved from http://proceedings.mlr.press/v81/buolamwini18a.html
[18] Jenna Burrell, Zoe Kahn, Anne Jonas, and Daniel Griffin. 2019. When Users Control the Algorithms: Values Expressed in Practices on Twitter. *Proceedings of the ACM on Human-Computer Interaction* 3, CSCW: 138:1-138:20. https://doi.org/10.1145/3359240
[19] Wendy Hui Kyong Chun. 2011. *Programmed visions: Software and memory*. mit Press.
[20] Lucas Colusso, Cynthia L Bennett, Pari Gabriel, and Daniela K Rosner. 2019. Design and Diversity? Speculations on what could go wrong. In *Proceedings of the 2019 on Designing Interactive Systems Conference*, 1405–1413.
[21] Danielle J. Corple and Jasmine R. Linabary. 2020. From data points to people: feminist situated ethics in online big data research. *International Journal of Social Research Methodology* 23, 2: 155–168. https://doi.org/10.1080/13645579.2019.1649832
[22] Nick Couldry and Ulises A Mejias. 2018. Data Colonialism: Rethinking Big Data's Relation to the Contemporary Subject. *Television & New Media* 20, 4: 336–349. https://doi.org/10.1177/1527476418796632
[23] Allan Dafoe. 2018. AI governance: a research agenda. *Governance of AI Program, Future of Humanity Institute, University of Oxford: Oxford, UK* 1442: 1443.
[24] Paul Dourish and Genevieve Bell. 2011. *Divining a Digital Future: Mess and Mythology in Ubiquitous Computing*. MIT Press, Cambridge.
[25] Paul Dourish and Scott D Mainwaring. 2012. Ubicomp's Colonial Impulse. In *Proceedings of the 2012 ACM Conference on Ubiquitous Computing* (UbiComp '12), 133–142. https://doi.org/10.1145/2370216.2370238
[26] Dylan Doyle-Burke and Jessie Smith. IBM, Microsoft, and Amazon Disavow Facial Recognition Technology: What Do You Need to Know? with Deb Raji. Retrieved October 6, 2020 from https://radicalai.podbean.com/e/ibm-microsoft-and-amazon-disavow-facial-recognition-technology-what-do-you-need-to-know-with-deb-raji/
[27] Paul N Edwards. 2003. Infrastructure and Modernity: Force, Time, and Social Organization in the History of Sociotechnical Systems. In *Modernity and Technology*, Thomas J Misa, Philip Brey and Andrew Feenberg (eds.). MIT Press, Cambridge, MA, 185–225.
[28] Upol Ehsan, Q. Vera Liao, Michael Muller, Mark O. Riedl, and Justin D. Weisz. 2021. Expanding Explainability: Towards Social Transparency in AI systems. In *Proceedings of the 2021 CHI Conference on Human Factors in Computing Systems*. Association for Computing Machinery, New York, NY, USA, 1–19. Retrieved June 15, 2021 from https://doi.org/10.1145/3411764.3445188
[29] Upol Ehsan and Mark O Riedl. 2019. On Design and Evaluation of Human-centered Explainable AI systems. In *Proceedings of Emerging Perspectives in Human-Centered Machine Learning: A Workshop at The ACM CHI Conference on Human Factors in Computing Systems*.
[30] Upol Ehsan and Mark O. Riedl. 2020. Human-Centered Explainable AI: Towards a Reflective Sociotechnical Approach. In *HCI International 2020 - Late Breaking Papers: Multimodality and Intelligence* (Lecture Notes in Computer Science), 449–466. https://doi.org/10.1007/978-3-030-60117-1_33
[31] Upol Ehsan, Pradyumna Tambwekar, Larry Chan, Brent Harrison, and Mark Riedl. 2019. Automated Rationale Generation: A Technique for Explainable AI and its Effects on Human Perceptions. In *Proceedings of the International Conference on Intelligence User Interfaces*, 263–274.
[32] Sl Eiin. Board of Intermediate and Secondary Education, Dhaka. 1. Retrieved from https://dhakaeducationboard.gov.bd/data/20200813101513549205.pdf
[33] Julia Elyachar. 2010. Phatic labor, infrastructure, and the question of empowerment in Cairo. *American Ethnologist* 37, 3: 452–464. https://doi.org/10.1111/j.1548-1425.2010.01265.x
[34] Motahhare Eslami, Aimee Rickman, Kristen Vaccaro, Amirhossein Aleyasen, Andy Vuong, Karrie Karahalios, Kevin Hamilton, and Christian Sandvig. 2015. "I always assumed that I wasn't really that close to [her]": Reasoning about Invisible Algorithms in News Feeds. In *Proceedings of the 33rd Annual ACM Conference on Human Factors in Computing Systems*, 153–162. https://doi.org/10.1145/2702123.2702556
[35] Virginia Eubanks. 2018. *Automating inequality: How high-tech tools profile, police, and punish the poor*. St. Martin's Press.
[36] Sina Fazelpour, Zachary C. Lipton, and David Danks. 2021. Algorithmic Fairness and the Situated Dynamics of Justice. *Canadian Journal of Philosophy*: 1–17. https://doi.org/10.1017/can.2021.24
[37] Batya Friedman, Peter H Kahn, and Alan Borning. 2008. Value sensitive design and information systems. *The handbook of information and computer ethics*: 69–101.
[38] Joan H Fujimura. 1991. On Methods, Ontologies, and Representation in the Sociology of Science: Where Do We Stand? In *Social Organization and Social Process: Essays in Honor of Anselm Strauss*, David Maines (ed.). Aldine de Gruyter, Hawthorne, NY, 207–248.
[39] Timnit Gebru, Jamie Morgenstern, Briana Vecchione, Jennifer Wortman Vaughan, Hanna Wallach, Hal Daumé III, and Kate Crawford. 2018. Datasheets for Datasets. In *Proceedings of the 5th Workshop on Fairness, Accountability, and Transparency in Machine Learning*. Retrieved July 6, 2020 from http://arxiv.org/abs/1803.09010
[40] Sucheta Ghoshal, Rishma Mendhekar, and Amy Bruckman. 2020. Toward a Grassroots Culture of Technology Practice. *Proceedings of the ACM on Human-Computer Interaction* 4, CSCW1: 054:1-054:28. https://doi.org/10.1145/3392862
[41] Stephen Graham and Simon Marvin. 2001. *Splintering Urbanism: Networked Infrastructures, Technological Mobilities and the Urban Condition*. Routledge, London.
[42] Ben Green and Salomé Viljoen. 2020. Algorithmic realism: expanding the boundaries of algorithmic thought. In *Proceedings of the 2020 Conference on Fairness, Accountability, and Transparency* (FAT* '20), 19–31. https://doi.org/10.1145/3351095.3372840
[43] Nina Grgic-Hlaca, Elissa M. Redmiles, Krishna P. Gummadi, and Adrian Weller. 2018. Human Perceptions of Fairness in Algorithmic Decision Making: A Case Study of Criminal Risk Prediction. In *Proceedings of the 2018 World Wide Web Conference* (WWW '18), 903–912. https://doi.org/10.1145/3178876.3186138
[44] Lelia Marie Hampton. 2021. Black Feminist Musings on Algorithmic Oppression. In *Proceedings of the 2021 ACM Conference on Fairness, Accountability, and Transparency* (FAccT '21), 1. https://doi.org/10.1145/3442188.3445929
[45] Donna Haraway. 1988. Situated knowledges: The science question in feminism and the privilege of partial perspective. *Feminist studies* 14, 3: 575–599.
[46] Rebecca Heilweil. 2021. Facebook is backing away from facial recognition. Meta isn't. *Vox*. Retrieved January 9, 2022 from https://www.vox.com/recode/22761598/facebook-facial-recognition-meta
[47] Alex Hern. 2020. Ofqual's A-level algorithm: why did it fail to make the grade? *The Guardian*. Retrieved January 14, 2022 from https://www.theguardian.com/education/2020/aug/21/ofqual-exams-algorithm-why-did-it-fail-make-grade-a-levels
[48] Ben Hutchinson, Andrew Smart, Alex Hanna, Emily Denton, Christina Greer, Oddur Kjartansson, Parker Barnes, and Margaret Mitchell. 2021. Towards Accountability for Machine Learning Datasets: Practices from Software Engineering and Infrastructure. In *Proceedings of the 2021 ACM Conference on Fairness, Accountability, and Transparency* (FAccT '21), 560–575. https://doi.org/10.1145/3442188.3445918
[49] Mizuko Ito. 2012. *Engineering play: A cultural history of children's software*. MIT Press.
[50] Abigail Z. Jacobs and Hanna Wallach. 2021. Measurement and Fairness. In *Proceedings of the 2021 ACM Conference on Fairness, Accountability, and Transparency*., 375–385. Retrieved December 25, 2019 from http://arxiv.org/abs/1912.05511
[51] Sheila Jasanoff (ed.). 2004. *States of knowledge: the co-production of science and social order*. Routledge, London; New York.
[52] Maximilian Kasy and Rediet Abebe. 2021. Fairness, Equality, and Power in Algorithmic Decision-Making. In *Proceedings of the 2021 ACM Conference on Fairness, Accountability, and Transparency* (FAccT '21), 576–586. https://doi.org/10.1145/3442188.3445919
[53] Michael Katell, Meg Young, Dharma Dailey, Bernease Herman, Vivian Guetler, Aaron Tam, Corinne Binz, Daniella Raz, and P M Krafft. 2020. Toward Situated Interventions for Algorithmic Equity: Lessons from the Field. 11.
[54] Helen Kennedy, Thomas Poell, and Jose van Dijck. 2015. Data and agency. *Big Data & Society* 2, 2: 1–7. https://doi.org/10.1177/2053951715621569
[55] Rob Kitchin and Martin Dodge. 2014. *Code/space: Software and everyday life*. Mit Press.
[56] Bernard Koch, Emily Denton, Alex Hanna, and Jacob Gates Foster. 2021. Reduced, Reused and Recycled: The Life of a Dataset in Machine Learning Research. *Proceedings of the Neural Information Processing Systems Track on Datasets and Benchmarks*. Retrieved December 13, 2021 from https://openreview.net/forum?id=zNQBIBKJRkd
[57] Vivian Lai and Chenhao Tan. 2019. On Human Predictions with Explanations and Predictions of Machine Learning Models: A Case Study on Deception Detection. In *Proceedings of the Conference on Fairness, Accountability, and Transparency* (FAT* '19), 29–38. https://doi.org/10.1145/3287560.3287590
[58] Martha Lampland and Susan Leigh Star. 2009. *Standards and their Stories: How Quantifying, Classifying, and Formalizing Practices shape Everyday Life*. Cornell University Press, Ithaca.

FAccT '22, June 21–24, 2022, Seoul, Republic of Korea    Upol Ehsan et al.[59] Brian Larkin. 2008. *Signal and Noise: Media, Infrastructure, and Urban Culture in Nigeria.* Duke University Press, Durham, NC.
[60] Charlotte P Lee, Paul Dourish, and Gloria Mark. 2006. The human infrastructure of cyberinfrastructure. In *Proceedings of the 2006 20th anniversary conference on Computer supported cooperative work*, 483–492.
[61] Min Kyung Lee. 2018. Understanding perception of algorithmic decisions: Fairness, trust, and emotion in response to algorithmic management. *Big Data & Society* 5, 1: 1–16. https://doi.org/10.1177/2053951718756684
[62] Min Kyung Lee, Daniel Kusbit, Anson Kahng, Ji Tae Kim, Xinran Yuan, Allissa Chan, Daniel See, Ritesh Noothigattu, Siheon Lee, Alexandros Psomas, and others. 2019. WeBuildAI: Participatory framework for algorithmic governance. *Proceedings of the ACM on Human-Computer Interaction* 3, CSCW: 1–35.
[63] Paul M. Leonardi. 2011. When Flexible Routines Meet Flexible Technologies: Affordance, Constraint, and the Imbrication of Human and Material Agencies. *MIS Quarterly* 35, 1: 147–167. https://doi.org/10.2307/23043493
[64] Paul M. Leonardi. 2013. Theoretical foundations for the study of sociomateriality. *Information and Organization* 23, 2: 59–76. https://doi.org/10.1016/j.infoandorg.2013.02.002
[65] Q Vera Liao, Daniel Gruen, and Sarah Miller. 2020. Questioning the AI: Informing Design Practices for Explainable AI User Experiences. In *Proceedings of the SIGCHI Conference on Human Factors in Computing Systems.*
[66] Duri Long and Brian Magerko. 2020. What is AI Literacy? Competencies and Design Considerations. In *Proceedings of the 2020 CHI Conference on Human Factors in Computing Systems* (CHI '20), 1–16. https://doi.org/10.1145/3313831.3376727
[67] Michael Lynch. 2016. Social Constructivism in Science and Technology Studies. *Human Studies* 39, 1: 101–112. https://doi.org/10.1007/s10746-016-9385-5
[68] Jonathan Magnusson. Improving Dark Pattern Literacy of End Users. 3.
[69] Jacob Metcalf, Emanuel Moss, and Danah Boyd. 2019. Owning Ethics: Corporate Logics, Silicon Valley, and the Institutionalization of Ethics. *Social Research: An International Quarterly* 86, 2: 449–476. Retrieved August 29, 2019 from https://muse.jhu.edu/article/732185
[70] Jacob Metcalf, Emanuel Moss, Elizabeth Anne Watkins, Ranjit Singh, and Madeleine Clare Elish. 2021. Algorithmic Impact Assessments and Accountability: The Co-construction of Impacts. In *Proceedings of the 2021 ACM Conference on Fairness, Accountability, and Transparency* (FAccT '21), 735–746. https://doi.org/10.1145/3442188.3445935
[71] Sabelo Mhlambi. 2020. From Rationality to Relationality: Ubuntu as an Ethical and Human Rights Framework for Artificial Intelligence Governance. Retrieved from https://carrcenter.hks.harvard.edu/publications/rationality-relationality-ubuntu-ethical-and-human-rights-framework-artificial
[72] Stefania Milan and Emiliano Treré. 2020. Big Data from the South(s): An Analytical Matrix to Investigate Data at the Margins. In *The Oxford Handbook of Sociology and Digital Media*, Deana A. Rohlinger and Sarah Sobieraj (eds.). Oxford University Press, Oxford.
[73] Margaret Mitchell, Simone Wu, Andrew Zaldivar, Parker Barnes, Lucy Vasserman, Ben Hutchinson, Elena Spitzer, Inioluwa Deborah Raji, and Timnit Gebru. 2019. Model Cards for Model Reporting. *Proceedings of the Conference on Fairness, Accountability, and Transparency - FAT\* '19*: 220–229. https://doi.org/10.1145/3287560.3287596
[74] Shira Mitchell, Eric Potash, Solon Barocas, Alexander D'Amour, and Kristian Lum. 2021. Algorithmic Fairness: Choices, Assumptions, and Definitions. *Annual Review of Statistics and Its Application* 8, 1: 141–163. https://doi.org/10.1146/annurev-statistics-042720-125902
[75] Matthew Moon. 2015. The Importance of AS-Level. Retrieved January 9, 2022 from https://www.cao.cam.ac.uk/behind-the-headlines/importance-of-aslevel
[76] Emanuel Moss, Elizabeth Anne Watkins, Ranjit Singh, Madeleine Clare Elish, and Jacob Metcalf. 2021. *Assembling Accountability Through Algorithmic Impact Assessment.* Data & Society Research Institute. Retrieved from http://datasociety.net/library/assembling-accountability/
[77] Arvind Narayanan. 2018. Translation tutorial: 21 fairness definitions and their politics. In *Proc. Conf. Fairness Accountability Transp., New York, USA.*
[78] Condé Nast. Everything that went wrong with the botched A-Levels algorithm. *Wired UK.* Retrieved January 14, 2022 from https://www.wired.co.uk/article/alevel-exam-algorithm
[79] Safiya Umoja Noble. 2018. *Algorithms of Oppression: How Search Engines Reinforce Racism.* NYU Press, New York.
[80] Ziad Obermeyer, Brian Powers, Christine Vogeli, and Sendhil Mullainathan. 2019. Dissecting racial bias in an algorithm used to manage the health of populations. *Science (New York, N.Y.)* 366, 6464: 447–453. https://doi.org/10.1126/science.aax2342
[81] Ofqual. The Office of Qualifications and Examinations Regulation (Ofqual) | About us. *GOV.UK.* Retrieved January 5, 2022 from https://www.gov.uk/government/organisations/ofqual/about
[82] Ofqual. Written statement from Chair of Ofqual to the Education Select Committee. *GOV.UK.* Retrieved October 28, 2021 from https://www.gov.uk/government/news/written-statement-from-chair-of-ofqual-to-the-education-select-committee
[83] W. J. Orlikowski. 2010. The sociomateriality of organisational life: considering technology in management research. *Cambridge Journal of Economics* 34, 1: 125–141. https://doi.org/10.1093/cje/bep058
[84] Jon Porter. 2020. UK ditches exam results generated by biased algorithm after student protests. *The Verge.* Retrieved January 14, 2022 from https://www.theverge.com/2020/8/17/21372045/uk-a-level-results-algorithm-biased-coronavirus-covid-19-pandemic-university-applications
[85] Manish Raghavan, Solon Barocas, Jon Kleinberg, and Karen Levy. 2020. Mitigating bias in algorithmic hiring: evaluating claims and practices. In *Proceedings of the 2020 Conference on Fairness, Accountability, and Transparency* (FAT\* '20), 469–481. https://doi.org/10.1145/3351095.3372828
[86] Inioluwa Deborah Raji, Andrew Smart, Rebecca N. White, Margaret Mitchell, Timnit Gebru, Ben Hutchinson, Jamila Smith-Loud, Daniel Theron, and Parker Barnes. 2020. Closing the AI accountability gap: defining an end-to-end framework for internal algorithmic auditing. In *Proceedings of the 2020 Conference on Fairness, Accountability, and Transparency*, 33–44.
[87] Bernhard Rieder. 2020. *Engines of order: A mechanology of algorithmic techniques.* Amsterdam University Press.
[88] Mary Beth Rosson and John M Carroll. 2009. Scenario based design. *Human-computer interaction. boca raton, FL*: 145–162.
[89] Christian Sandvig, Kevin Hamilton, Karrie Karahalios, and Cedric Langbort. 2014. An Algorithmic Audit. In *An Algorithmic Audit*, Gangadharan (ed.). Open Technology Institute, available at: https://www.newamerica.org/downloads/OTI-Dataan-Discrimination-FINAL-small.pdf (accessed 3 July 2015), Data and Discrimination: Collected Essays. Retrieved October 16, 2015 from http://www.kevinhamilton.org/share/papers/OTI-Data-an-Discrimination-FINAL-small.pdf
[90] Aaron Sankin, Dhruv Mehrotra for Gizmodo, Surya Mattu, and Annie Gilbertson. 2021. Crime Prediction Software Promised to Be Free of Biases. New Data Shows It Perpetuates Them. *The Markup.* Retrieved December 13, 2021 from https://themarkup.org/prediction-bias/2021/12/02/crime-prediction-software-promised-to-be-free-of-biases-new-data-shows-it-perpetuates-them
[91] Nick Seaver. 2017. Algorithms as culture: Some tactics for the ethnography of algorithmic systems. *Big Data & Society* 4, 2: 2053951717738104. https://doi.org/10.1177/2053951717738104
[92] Andrew D. Selbst, Danah Boyd, Sorelle A. Friedler, Suresh Venkatasubramanian, and Janet Vertesi. 2019. Fairness and Abstraction in Sociotechnical Systems. In *Proceedings of the Conference on Fairness, Accountability, and Transparency - FAT\* '19*, 59–68. https://doi.org/10.1145/3287560.3287598
[93] Mark Sendak, Madeleine Clare Elish, Michael Gao, Joseph Futoma, William Ratliff, Marshall Nichols, Armando Bedoya, Suresh Balu, and Cara O'Brien. 2020. "The human body is a black box": supporting clinical decision-making with deep learning. In *Proceedings of the 2020 Conference on Fairness, Accountability, and Transparency* (FAT\* '20), 99–109. https://doi.org/10.1145/3351095.3372827
[94] Phoebe Sengers, Kirsten Boehner, Shay David, and Joseph'Jofish' Kaye. 2005. Reflective design. In *Proceedings of the 4th decennial conference on Critical computing: between sense and sensibility*, 49–58.
[95] Hong Shen, Alicia DeVos, Motahhare Eslami, and Kenneth Holstein. 2021. Everyday Algorithm Auditing: Understanding the Power of Everyday Users in Surfacing Harmful Algorithmic Behaviors. *Proceedings of the ACM on Human-Computer Interaction* 5, CSCW2: 433:1-433:29. https://doi.org/10.1145/3479577
[96] Katie Shilton, Emanuel Moss, Sarah A. Gilbert, Matthew J. Bietz, Casey Fiesler, Jacob Metcalf, Jessica Vitak, and Michael Zimmer. 2021. Excavating awareness and power in data science: A manifesto for trustworthy pervasive data research. *Big Data & Society* 8, 2: 20539517211040760. https://doi.org/10.1177/20539517211040759
[97] Ben Shneiderman. 2016. Opinion: The dangers of faulty, biased, or malicious algorithms requires independent oversight. *Proceedings of the National Academy of Sciences* 113, 48: 13538–13540. https://doi.org/10.1073/pnas.1618211113
[98] AbdouMaliq Simone. 2004. People as infrastructure: Intersecting fragments in Johannesburg. *Public culture* 16, 3: 407–429.
[99] Ranjit Singh. 2020. Study the Imbrication: A Methodological Maxim to Follow the Multiple Lives of Data. In *Lives of Data: Essays on Computational Culture in India*, Sandeep Mertia (ed.). Institute of Network Cultures, Amsterdam, 51–59. Retrieved from https://networkcultures.org/wp-content/uploads/2021/02/Lives-of-Data-.pdf
[100] Ranjit Singh and Steven Jackson. 2021. Seeing Like an Infrastructure: Low-resolution Citizens and the Aadhaar Identification Project. *Proceedings of the ACM on Human-Computer Interaction* 5, CSCW2: 315:1-315:26. https://doi.org/10.1145/3476056
[101] Ranjit Singh and Steven J. Jackson. 2017. From Margins to Seams: Imbrication, Inclusion, and Torque in the Aadhaar Identification Project. In *Proceedings of the 2017 CHI Conference on Human Factors in Computing Systems.*
[102] Susan Leigh Star. 1991. Power, Technology, and the Phenomenology of Conventions: On being Allergic to Onions. In *A Sociology of Monsters: Essays on Power, Technology, and Domination*, John Law (ed.). Routledge, London, 26–56.
[103] Susan Leigh Star. 1999. The Ethnography of Infrastructure. *American Behavioral Scientist* 43, 3: 377–391. https://doi.org/10/b7hh4b

[104] Susan Leigh Star and Karen Ruhleder. 1996. Steps Toward an Ecology of Infrastructure: Design and Access for Large Information Spaces. *Information Systems Research* 7, 1: 111–134. https://doi.org/10/fdqbw8

[105] John M Staudenmaier. 1985. *Technology's Storytellers: Reweaving the Human Fabric*. The Society for the History of Technology and the MIT Press, Cambridge, MA.

[106] Lucy Suchman and Lucy A Suchman. 2007. *Human-machine reconfigurations: Plans and situated actions*. Cambridge university press.

[107] Janet Vertesi. 2014. Seamful Spaces: Heterogeneous Infrastructures in Interaction. *Science, Technology & Human Values* 39, 2: 264–284. Retrieved from http://sth.sagepub.com/content/39/2/264.abstract

[108] Salome Viljoen. 2020. *A Relational Theory of Data Governance*. Social Science Research Network, Rochester, NY. https://doi.org/10.2139/ssrn.3727562

[109] Elizabeth Anne Watkins, Emanuel Moss, Jacob Metcalf, Ranjit Singh, and Madeleine Clare Elish. 2021. Governing Algorithmic Systems with Impact Assessments: Six Observations. In *Proceedings of the 2021 AAAI/ACM Conference on AI, Ethics, and Society* (AIES '21), 1010–1022. https://doi.org/10.1145/3461702.3462580

[110] Allison Woodruff, Sarah E. Fox, Steven Rousso-Schindler, and Jeffrey Warshaw. 2018. A Qualitative Exploration of Perceptions of Algorithmic Fairness. In *Proceedings of the 2018 CHI Conference on Human Factors in Computing Systems*. Association for Computing Machinery, New York, NY, USA, 1–14. Retrieved December 8, 2021 from https://doi.org/10.1145/3173574.3174230

[111] Malte Ziewitz and Ranjit Singh. 2021. Critical companionship: Some sensibilities for studying the lived experience of data subjects. *Big Data & Society* 8, 2: 1–13. https://doi.org/10.1177/20539517211061122

[112] 2020. All English medium schools must have government registration: Ministry. *The Business Standard*. Retrieved January 9, 2022 from https://www.tbsnews.net/bangladesh/education/all-english-medium-schools-must-have-government-registration-ministry-113293

[113] 2020. "F**k the algorithm"?: What the world can learn from the UK's A-level grading fiasco. *Impact of Social Sciences*. Retrieved January 14, 2022 from https://blogs.lse.ac.uk/impactofsocialsciences/2020/08/26/fk-the-algorithm-what-the-world-can-learn-from-the-uks-a-level-grading-fiasco/

[114] 2021. Facebook to Shut Down Use of Facial Recognition Technology. *Bloomberg.com*. Retrieved January 9, 2022 from https://www.bloomberg.com/news/articles/2021-11-02/facebook-to-shut-down-use-of-facial-recognition-technology

[115] Where are Cambridge International AS & A Levels accepted and recognised? *What can we help you with?* Retrieved January 9, 2022 from https://help.cambridgeinternational.org/hc/en-gb/articles/115004302785-Where-are-Cambridge-International-AS-A-Levels-accepted-and-recognised-

[116] Guide to AS and A level results for England, 2019. *GOV.UK*. Retrieved January 14, 2022 from https://www.gov.uk/government/news/guide-to-as-and-a-level-results-for-england-2019

[117] Regulating GCSEs, AS and A levels: guide for schools and colleges. *GOV.UK*. Retrieved January 9, 2022 from https://www.gov.uk/guidance/regulating-gcses-as-and-a-levels-guide-for-schools-and-colleges

[118] IGCSE | British Council. Retrieved January 9, 2022 from https://www.britishcouncil.org.bd/en/exam/igcse-school

[119] Exam registration for all candidates | British Council. Retrieved January 9, 2022 from https://www.britishcouncil.org.bd/en/exam/igcse-school/register/private-candidates

[120] Facebook reaches $550 million settlement in facial recognition lawsuit. *NBC News*. Retrieved January 9, 2022 from https://www.nbcnews.com/tech/tech-news/facebook-reaches-550-million-settlement-facial-recognition-lawsuit-n1126191